\documentclass[aps,pra,twocolumn]{revtex4}

\usepackage{amsmath}
\usepackage{amssymb}
\usepackage{mathrsfs}
\usepackage{verbatim}
\usepackage{epsfig}
\usepackage{color}
\usepackage[bookmarks]{hyperref}
\usepackage{graphicx}

\graphicspath{{.}{./EPS/}}

\newcommand* {\bra}[1]{\ensuremath{\langle {#1} |}}
\newcommand* {\ket}[1]{\ensuremath{| {#1} \rangle}}

\newcommand{\ketbra}[2]{|{#1}\rangle\langle{#2}|}
\newcommand{\braket}[2]{\langle{#1}|{#2}\rangle}

\newcommand{\hbrho}{\boldsymbol{\rho}}

\begin{document}
\title{Conditions for entanglement purification with general two-qubit states}

\author{Juan Mauricio Torres}
\email{mauricio.torres@physik.tu-darmstadt.de}
\affiliation{Institut f\"{u}r Angewandte Physik, Technische Universit\"{a}t Darmstadt, D-64289 Darmstadt, Germany}
\author{J\'ozsef Zsolt Bern\'ad}
\email{Zsolt.Bernad@physik.tu-darmstadt.de}
\affiliation{Institut f\"{u}r Angewandte Physik, Technische Universit\"{a}t Darmstadt, D-64289 Darmstadt, Germany}

\date{\today}

\begin{abstract}
We present the convergence study of a recurrence entanglement purification protocol
using arbitrary two-qubit initial states.
The protocol is based on a rank two projector in the Bell basis which serves
as a two-qubit operation replacing the usual controlled-NOT gate. 
We show that the whole space of two-qubit density matrices is mapped onto an invariant subspace characterized
by seven real parameters. 
By analyzing this type of density matrices 
we are able to find general conditions
for entanglement purification in the form of two inequalities between pairs of diagonal elements and pairs of coherences.
We show that purifiable initial states do not necessary require a fidelity larger than one half with respect to any
maximally entangled pure state.
Furthermore, we find a family of states parametrized by their concurrence 
that can be perfectly converted into a Bell state in just one step of the protocol with probability proportional
to the square of the concurrence.
\end{abstract}

\maketitle

\section{Introduction}

Entanglement purification protocols \cite{Bennett1,Bennett2,Deutsch} generate Bell states 
from an ensemble of noisy entangled states.
They have found application in many areas of quantum information theory \cite{DürBriegel} and  
prominently in proposed quantum communication technologies \cite{Briegel98}.
From a mathematical perspective, the effect of these protocols can be described as a non-linear 
map of the parameter defining the input density matrices.
The first of the so-called recurrence protocols \cite{DürBriegel} was introduced by Bennett 
and collaborators \cite{Bennett1}. This protocol relies on 
Werner states \cite{Werner} which are described by one parameter and therefore 
the convergence of the corresponding map \cite{Bennett1} can be studied in a simple way.
A more efficient protocol was introduced by Deutsch et al. \cite{Deutsch}
based on Bell diagonal states involving a three dimensional parameter space. 
The convergence analysis is more intricate in this case and was studied separately in detail by C. 
Macchiavello \cite{Macchiavello}. 

There are two essential steps in recurrence entanglement purification protocols \cite{DürBriegel}, which are reconsidered 
in this work: First, the application of random local unitary rotations transforming any initial 
density matrix into a Werner \cite{Deutsch} or a 
Bell-diagonal state \cite{Bennett2}. Second, the application of bilateral controlled-NOT gates on the spatially separated quantum systems. The latter step has already been modified in our proposed purification protocol \cite{Bernad}
based on Refs. \cite{Bennett1,Deutsch}.
The modification was motivated by the design of a multiphoton-assisted quantum repeater  and
instead of controlled-NOT gates it is based on a rank two projection in the Bell basis 
which can be implemented in a one-atom maser setup \cite{Haroche}. We have studied in Ref. \cite{Bernad} the performance
of the modified protocol using Werner and other Bell-diagonal states. 
A natural question arises whether this protocol can still work for arbitrary input states making 
the step of random local unitary rotations (``twirling'') superfluous. This problem holds  great significance, because  random operations 
can waste important entanglement and their implementation is not straightforward. 
Therefore, a solution to this problem could offer a less intricate purification protocol 
and a better compatibility with certain experimental settings.

In this paper we show that our protocol, first introduced in Ref. \cite{Bernad}, can exploit the entanglement
of arbitrary initial states without using random unitary operations. 
After one iteration of the protocol, any input state is mapped onto a density matrix described by seven real parameters.
This form is invariant under further iterations of the protocol. 
We exploit this feature to analyze the convergence properties for any input state of this form. 
Based on this analysis, we generalize to the case of arbitrary two-qubit states and find 
conditions that allow purification of Bell states.
We demonstrate that in contrast to previous schemes \cite{Bennett1,Deutsch}, a purifiable state does not require
an initial fidelity larger than one half with respect to any Bell state. In particular, we find a class of states
with overlap less than one half with any maximally entangled pure state that can be 
purified in just one step of the protocol. 

The paper is organized as follows. In Sec. \ref{PurifyP} we reintroduce and explain the steps of our protocol. 
In Sec. \ref{Purify7} we study the set of density matrices characterized by seven real parameters
and derive the conditions for a successful entanglement purification. In Sec. \ref{Purify15} we extend the analysis to density matrices characterized by fifteen real parameters. 
Examples which obey the newly found conditions are presented in Sec. \ref{Examples}. In Appendix \ref{App} 
we present the stability analysis of  the fixed points for the seven dimensional parameter space.

\section{Entanglement purification protocol}
\label{PurifyP}
In this section we review the entanglement purification protocol introduced in Ref. \cite{Bernad}
which is based on the protocols of Refs. \cite{Bennett1,Deutsch}.
Let us consider as initial condition the product state of two qubit pairs 
\begin{equation}
  \hbrho=\rho^{A_1,B_1}\rho^{A_2,B_2}.
\label{rhoAB4}
\end{equation}
Both pairs are assumed to start in the same state $\rho$ with certain degree of entanglement
and their qubit components to be in distant locations labelled by $A$ and $B$. The aim of an entanglement purification protocol is to 
trade two pairs for one pair with larger degree of entanglement using only local operations in
laboratories $A$ and $B$ and classical communication between them. This can be achieved by increasing the 
fidelity with respect to any 
of the Bell states
\begin{align}
  \ket{\Psi^\pm}=\tfrac{1}{\sqrt2}\left(\ket{01}\pm\ket{10}\right), \,
  \ket{\Phi^\pm}=\tfrac{1}{\sqrt2}\left(\ket{00}\pm \ket{11}\right).
  \label{Bellstates}
\end{align}

The purification protocol that we consider in this work consists of the following steps: \\
(I) The two-qubit quantum operation $M$ is applied locally in locations $A$ and $B$, where
\begin{equation}
M=\ket{\Psi^{-}}\bra{\Psi^{-}}+\ket{\Phi^{-}}\bra{\Phi^{-}}. 
\end{equation}
After a successfully applied quantum operation the four qubit system attains the state  
\begin{equation}
\hbrho'=
\frac{ \Pi \hbrho \Pi^\dagger }{\mathrm{Tr} 
\left\{\Pi^\dagger \Pi \hbrho\right\}}
,\quad  \Pi=M^{A_1,A_2}  M^{B_1,B_2}.
\label{maprho}
\end{equation}
(II)  One of the pairs is then locally measured in the computational basis, say pair $(A_2,B_2)$.
The choice of the measured pair is unimportant as the operation is symmetric and the initial qubit 
pairs are identical. There are four possible states in which one can find pair $(A_2,B_2)$: 
$\ket{jk}^{A_2,B_2}\equiv\ket{j}^{A_2}\ket{k}^{B_2}$ with $j,k\in\{0,1\}$. A successful measurement
of one of the states $\ket{jk}^{A_2,B_2}$ results in
the two qubit state
\begin{align}
  \tilde\rho^{A_1,B_1}=\mathrm{Tr}_{A2,B2}\left\{\ketbra{jk}{jk}^{A_2,B_2} \hbrho'\right\}.
\label{}
\end{align}
(III) Depending on the measurement results, the quantum gate 
$V_j^{A_1} V_{k+1}^{B_1}$ is applied  to the remaining qubit pair, where
\begin{align}
V_j=
\left(\ket{1}\bra{1}+i\ket{0}\bra{0}\right){\sigma_x}^j
\label{}
\end{align}
with the Pauli operator $\sigma_x=\ket{1}\bra{0}+\ket{0}\bra{1}$. 
The final two-qubit state is  then given by
\begin{align}
  \rho'^{A_1,B_1}=
  \left(V_j^{A_1} V_{k+1}^{B_1}\right)
  \tilde\rho^{A_1,B_1}
  \left(V_j^{A_1} V_{k+1}^{B_1}\right)^\dagger.
  \label{}
\end{align}
We remark again that there is a free choice of the qubit pair to be measured as our quantum
operation $M$ is symmetric in contrast to the controlled-NOT gate used in the seminal protocols
\cite{Bennett1,Deutsch}. As the entangled pairs are assumed to start in the same state $\rho$,
in what follows we will omit the labels and study the two qubit map $\rho\to\rho'$.

Now that we have explained the key elements of the protocol, we will study the resulting
map between initial and output states of the qubit pairs. Consider 
a general two-qubit density matrix that can be written as
\begin{eqnarray}
  \rho&=&\sum^4_{i,j=1} r_{ij} \ket{i} \bra{j}, \quad r_{j}\equiv r_{jj} 
 \label{representation2}
\end{eqnarray}
where we have chosen to work in the Bell basis labeled in the following way
\begin{align}
 \ket{1}\equiv\ket{\Psi^-},\,\,\ket{2}\equiv\ket{\Phi^-},\,\,
 \ket{3}\equiv\ket{\Phi^+}, \,\, \ket{4}\equiv\ket{\Psi^+}. 
  \label{basis}
\end{align}
Due to the conditions $\mathrm{Tr}\{\rho\}=1$ and $\rho^\dagger=\rho$ we have the following relations:
\begin{eqnarray}
 r_1+r_2+r_3+r_4=1,\quad r_{ij}=\left(r_{ji}\right)^*. \nonumber
\end{eqnarray}
The probabilistic part of this protocol takes place in step (I), i.e., in  the application of the bilateral two-qubit
quantum operation $M$. The success probability of the quantum operation is $N/2$, with 
\begin{equation}
N=(r_{1}+r_{2})^2+(r_{3}+r_{4})^2-(r_{12}+r_{21})^2-(r_{34}+r_{43})^2.
\end{equation}
Provided both implementations of $M$ are successful, the steps (II) and (III) can be given for granted
as they involve unitary gates and none of the qubit measurements is discarded. 
The resulting two-qubit density matrix $\rho'$ in the same basis 
as \eqref{representation2} has the non-zero entries
\begin{align}
 r'_1&=\frac{r^2_{1}+r^2_{2}-r^2_{12}-r^2_{21}}{N}, \quad r'_2=2\frac{r_{3}r_{4}-|r_{34}|^2}{N}, \nonumber \\
 r'_4&=\frac{r^2_{3}+r^2_{4}-r^2_{34}-r^2_{43}}{N},
 \quad
 r'_3=2\frac{r_{1}r_{2}-|r_{12}|^2}{N}, 
 \label{map2}  \\
  r_{14}'&=\frac{r^2_{14}+r^2_{23}-r^2_{13}-r^2_{24}}{N},  
  \quad r'_{23}=2\frac{r^*_{23}r^*_{14}-r^*_{13}r^*_{24}}{N}.
 \nonumber
\end{align}
Together with 
$r_{41}'=(r_{14}')^\ast$ and $r_{32}'=(r_{23}')^\ast$, these are the only non-vanishing elements
of $\rho'$ after a single iteration of the protocol. The constant $N$ plays here the role of a normalization factor and
the resulting density matrix has eight vanishing entries. Therefore, only seven parameters characterize
these type of states, in contrast to the fifteen parameters needed to describe a general two-qubit 
state. This simplification motivates the study of the purification protocol for density matrices described by the seven primed parameters of Eq. \eqref{map2}. 

It is worth to note that the protocol in Ref. \cite{Deutsch} is also able to reduce the fifteen dimensional parameter space to a seven dimensional one. This happens
technically in a rather different way as in our protocol. After measuring the target pair during the procedure, one keeps the control pair only if the target pair was found in the state
$\ket{11}$. The result can be obtained by a straightforward application of the protocol in Ref.  \cite{Deutsch} on general two-qubit density matrices and up to our knowledge it has never been
reported. 

\section{Entanglement purification of $X$ states in the Bell basis}
\label{Purify7}
In this section we concentrate our study on initial two-qubit states described by seven free parameters, which have been 
delineated in Eq. \eqref{map2}.
In the Bell basis \eqref{basis} they have the  matrix representation 
\begin{equation}
  \rho=\left(
  \begin{array}{cccc}
    r_1 & 0 & 0 & r_{1_4}\\
    0 & r_2 & r_{23} & 0\\
    0 & r_{32} & r_3 & 0 \\
    r_{41} & 0 & 0 & r_4
  \end{array}
  \right),
  \label{representation}
\end{equation}
and therefore we refer to them them as $X$ states.
The diagonal elements  $r_1,r_2,r_3,r_4 \in [0,1]$ and the off-diagonal 
elements $r_{23},\,r_{14} \in \mathbb{C}$
fulfil
the conditions $r_1+r_2+r_3+r_4=1$, $r_{32}=r_{23}^\ast$ and $r_{41}=r_{14}^\ast$.
The four eigenvalues can be represented as the two pairs $\lambda_{23}^\pm$ and $\lambda_{14}^\pm$ which  can be obtained 
from the quadratic formula
\begin{eqnarray}
  \lambda_{jk}^\pm&=&\frac{r_j+r_k\pm\sqrt{(r_j-r_k)^2+4|r_{jk}|^2}}{2}.
 \label{eigenvalues}
\end{eqnarray}
As the eigenvalues of $\rho$ have to be real non-negative numbers, the radical in Eq. \eqref{eigenvalues}
has to be positive, and this imposes the following restriction to the absolute value of the coherences 
\begin{align}
  |r_{23}|\le\sqrt{r_2r_3},\quad
  |r_{14}|\le\sqrt{r_1r_4}.
  \label{offcondition}
\end{align}

In the case of input states in the form of Eq. \eqref{representation}, 
the output state $\rho'$ remains in the same form and according to \eqref{map2} with the new coefficients:
\begin{align}
 r'_1&=\frac{r^2_1+r^2_2}{N}, \quad r'_2=2\frac{r_3 r_4}{N}, 
 \quad r_{14}'=\frac{r_{14}^2+r_{23}^2}{N}, 
 \nonumber \\
 r'_4&=\frac{r^2_3+r^2_4}{N}, \quad
 r'_3=2\frac{r_1r_2}{N}, 
 \quad \, r'_{23}=2\frac{r^*_{23} r^*_{14}}{N}, \
 \label{map} 
\end{align}
and the normalization factor 
\begin{align}
N=
(r_1+r_2)^2+(1-r_1-r_2)^2.
  \label{normalization}
\end{align}
In order to find the convergence properties of this map, we will
first  study the behaviour of the off-diagonal elements  after one iteration
by considering the norm of the pair $(r_{14},r_{23})$.
It turns out that this quantity does not increase, but  typically decreases after iteration of the map. 
This  
can be noted by analyzing the ratio
\begin{eqnarray}
  \frac{|r'_{14}|+|r'_{23}|}{|r_{14}|+|r_{23}|}\le \frac{|r_{14}|+|r_{23}|}{N}\le 1,
 \label{normq}
\end{eqnarray}
where we have used the normalization factor in Eq. \eqref{normalization}.
The first inequality can be obtained by direct evaluation of $r_{14}'$ and $r_{23}'$ using Eq. \eqref{map}, and by
taking into account the triangle inequality
$|r_{14}^2+r_{23}^2| \leqslant |r_{14}|^2+|r_{23}|^2$. The second inequality in Eq. \eqref{normq} follows from
considering the ratio between the maximum value of $|r_{14}|+|r_{23}|$ and the minimum value of $N$, both of which are $1/2$.
To show this, first note that according to Eq. \eqref{normalization}, $N$  attains its minimum value 
$1/2$ when $r_1+r_2=1/2$. 
Secondly, we take into account the conditions in \eqref{offcondition}
that combined with the normalization condition for $\rho$ and the
fact that for any pair of real numbers $2x y\le x^2+y^2$, implies the relation $ |r_{14}|+|r_{23}| \leqslant 1/2$ 
between the coherences. 
Taking the equality case in \eqref{offcondition} one can note that 
the equality in Eq. \eqref{normq} is achieved only 
with the initial conditions
\begin{eqnarray}
  |r_{14}|&=&r_1=r_4=0.5, 
  \quad |r_{23}|=r_2=r_3=0, 
 \nonumber \\
 |r_{14}|&=&r_1=r_4=0,\quad 
 |r_{23}|=r_2=r_3=0.5,  \label{offspec}\\
 |r_{14}|&=&|r_{23}|=r_1=r_2=r_3=r_4=0.25. \nonumber
\end{eqnarray}
Only in these three cases the norm of the coherences remains constant after
repeated iteration of the map in \eqref{map}. For all other values of the coherences,
the ratio in Eq.\eqref{normq} is strictly less than one
and therefore $r_{14}$ and $r_{23}$ tend to zero by repeated iterations of the map \eqref{map}.
We remark that in all these three cases, all the diagonal values are less that one half.

Now we turn our attention to the diagonal elements. 
The aim is to find the initial conditions that after iteration of the map tend to one of the Bell states.
The protocol we have presented here  
is not the same as the  protocol in Ref. \cite{Deutsch}, however the map \eqref{map} for the diagonal 
elements $r_1,r_2,r_3,r_4$ has a similar structure. 
Therefore, we follow the idea of the proof in Ref. \cite{Macchiavello}, 
where Bell diagonal density matrices were studied.
The same treatment is justified in our case as the diagonal elements in Eq. \eqref{map} 
do not depend on the off-diagonal ones. 

We start this discussion by noting that $r_1'\ge r_3'$ and $r_4'\ge r_2'$ which indicates that 
neither $r_2$ or $r_3$ can reach unit value, actually not even above $1/2$. With this fact we
recognize that only $r_1$ or $r_4$ can potentially increase their value to unity after repeated
iterations of the map \eqref{map}. Noting the
symmetry of the map under interchange of $r_1\leftrightarrow r_4$ and $r_2\leftrightarrow r_3$, allows us to consider 
only the maximization of one, say $r_1$.  Indeed the case $r_1=1$
and all other parameters equal to zero is a fixed point of the map \eqref{map}. Now, let us 
rewrite the value of $r_1'$ as
\begin{equation*}
  r_1'=\frac{r_1^2+r_2^2}{2(r_1^2+r_2^2)-F(r_1,r_2)},
\end{equation*}
were we have introduced the quadratic form
\begin{equation}
 F(r_1,r_2)=(2r_1-1)(1-2r_2).
\end{equation}
Whenever $F(r_1,r_2)$ is positive, $r_1'$ attains values larger than $1/2$.
This actually happens whenever $r_1$ or $r_2$ are larger than $1/2$ and in this case
the function $F$ increases after  iteration of the map \eqref{map}, 
as can be noted from the  relation
\begin{align}
\frac{ F(r_1',r_2')}{ F(r_1,r_2)}=
\frac{N-4r_3r_4}{N^2}
=\frac{(1-x)^2+y^2}{[(1-x)^2+x^2]^2}\ge 1,
  \label{supercondition}
\end{align}
with $x=r_3+r_4$ and $y=r_4-r_3$.
The inequality follows from the fact that $1-x\ge(1-x)^2+x^2$ for $x\in [0,1/2)$, 
which holds true if $r_1>1/2$ or $r_2>1/2$. 
The function $F(r_1,r_2)$ increases its value after application of the map \eqref{map} when the strict inequality
is met. In this case, as $F(r_1,r_2)$ is a monotonic function of $r_1$ and $r_2$, iteration of the map
converges to the fixed point with $r_1=1$ which is the only fixed point with the condition $r_1>1/2$.
The equality in \eqref{supercondition}  happens only in the case $r_3=r_4=0$, but
after each iteration of the map \eqref{map} with initial values 
$r_3=r_4=0$, the values of $r_3$ and $r_4$ cannot be simultaneously zero and thus $F(r''_1,r''_2)>F(r'_1,r'_2)$ unless $r_2=0$
but that would be the case of the fixed point with $r_1=1$. 
This  proves that the  condition to purify to a Bell state $\ket{\Psi^-}$, i.e. $r_1=1$, can be written as
\begin{equation}
 (2r_1-1)(1-2r_2)>0
  \label{conditionr1}
\end{equation}
In complete analogy due to the symmetry $r_1\leftrightarrow r_4$ and $r_2\leftrightarrow r_3$, one can find the condition
to purify to Bell state $\ket{\Psi^+}$, i.e. $r_1=1$, can be written as
\begin{equation}
 (2r_4-1)(1-2r_3)>0.
  \label{conditionr4}
\end{equation}

\begin{figure}[t]
 \includegraphics[width=.4\textwidth]{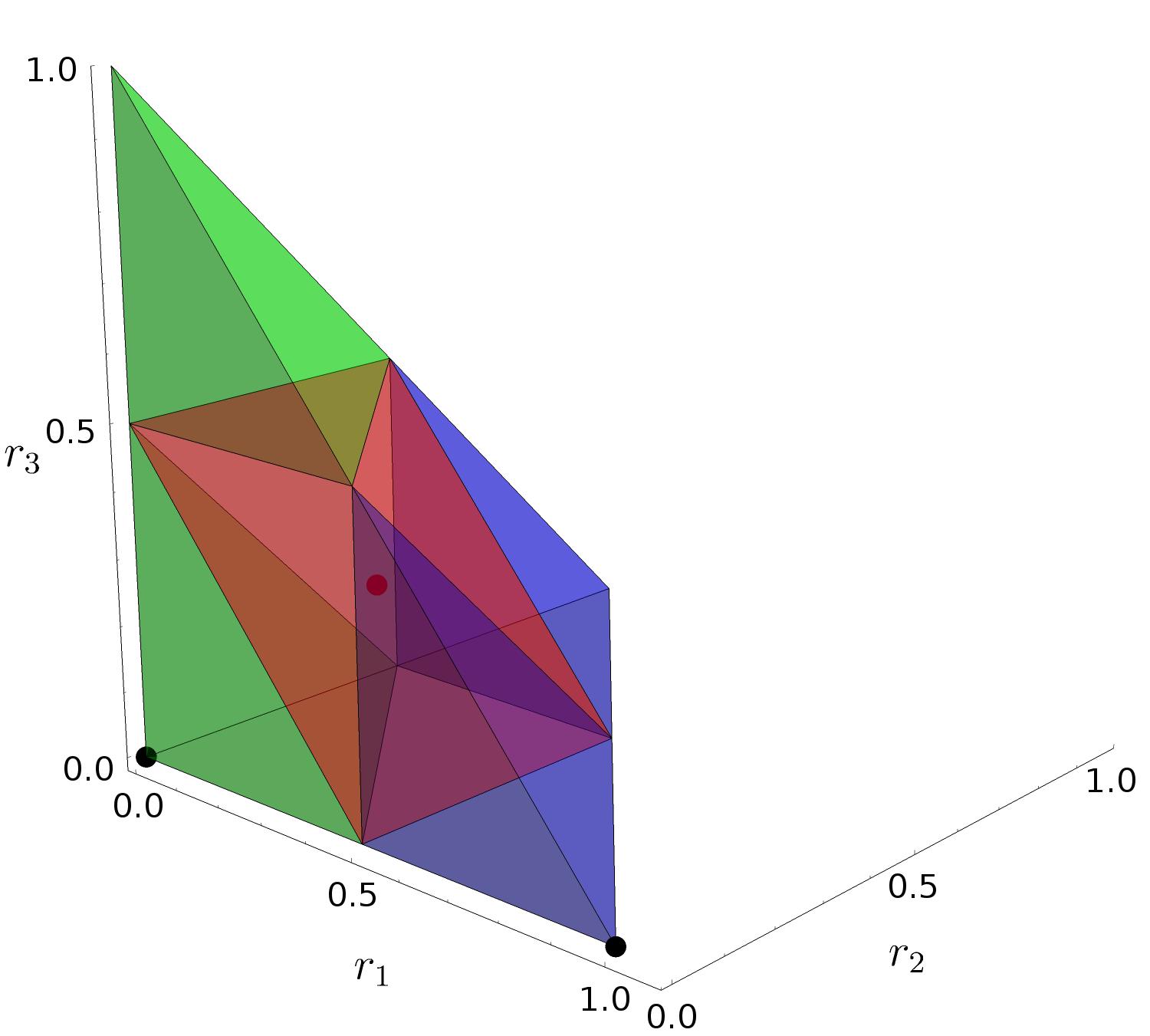}
 \caption{\label{space7}
 Three dimensional space of the independent diagonal elements $(r_1,r_2,r_3)$. The other parameter
 $r_4$ is obtained form the trace condition ${\rm Tr}\rho=1$. In blue: the region defined by condition \eqref{conditionr1} whose points lead to the fixed point $r_1=1$.
 In green: the region defined by condition \eqref{conditionr4} that lead to the fixed point $r_4=1$. In red: all other
 conditions that do not lead to an entangled state. The black dots are the stable fixed points 
 that correspond to Bell states, whereas the red one is the fixed point representing the totally mixed state.}
\end{figure}

These inequalities define four purifiable regions in the three dimensional space of the diagonal parameters shown in Fig. \ref{space7}. These findings are similar to 
Ref. \cite{Macchiavello}, where purifications of the states $\ket{\Phi^+}$ and $\ket{\Psi^+}$ have been proven.
Here, our proof is extended taking into account off-diagonal elements of $X$ states subject to our protocol. We have shown in Eqs. \eqref{normq} and \eqref{offspec} that the absolute value of the 
off-diagonal parameters decreases in the regions subject to the conditions \eqref{conditionr1} and \eqref{conditionr4}. Therefore the regions depicted in blue and green in Fig. \ref{space7} have one stable attractor point. 
In Appendix \ref{App} we evaluate numerically the stability of the fixed points. 

\section{Conditions for general two-qubit density matrices}
\label{Purify15}

In this section we extend our analysis to the case of general two-qubit density matrices in Eq. \eqref{representation2}.
In this case and after one iteration one obtains the density matrix of an $X$ state (see Eq. \eqref{representation}). Therefore,
the most important question is to define the conditions in which after the first iteration of the map 
\eqref{map2} one obtains any of the four $r_i'$'s larger than $1/2$.
First we note that one can never achieve $r_2'>1/2$ or $r_3'>1/2$ after one iteration. This due to 
the fact that  $r_1'\ge r_3'$ and $r_4'\ge r_2'$, which is noted from \eqref{map2}
by comparing their numerators and considering that: $x^2+y^2 \ge 2xy$ for any $x,y\in {\mathbb R}$ and 
because $z^2+z^{\ast 2}\le 2|z|^2$ for any $z \in {\mathbb C}$. 
This means that we only have to find those conditions that
leave $r_1'>1/2$ or $r_4'>1/2$. For symmetry reasons it suffices to focus in only one, we choose $r_1$.
Let us introduce the following abbreviations 
\begin{align}
 a&=(r_1^2+r_2^2-r_{12}^2-r_{21}^2), \\
 b&=(2r_1-1)(1-2r_2)-(r_{12}-r_{21})^2+(r_{34}+r_{43})^2, \nonumber
  \label{}
\end{align}
which relate to the normalization factor as $N=2a-b$.
This form of rewriting $N$ allows to identify the fidelity with respect to $\ket{\Psi^-}$
after one step of the map in Eq. \eqref{map2}, namely
\begin{equation}
r_1'=a/(2a-b).
\end{equation}
This expression is larger than $1/2$ whenever $b>0$ or equivalently when the following inequality is met
\begin{align}
  (2r_1-1)(1-2r_2)>-
  (2{\rm Im [r_{12}]})^2-(2{\rm Re}[r_{34}])^2.
  \label{genconditionr1}
\end{align}
This condition generalizes \eqref{conditionr1} as any state fulfilling it is transformed by the map into
an $X$ state of the form of \eqref{representation}, but  with the coefficient $r_1'>1/2$.
In Sec. \ref{Purify7} we proved that these type of states can be  purified. 
Because the right-hand side of Eq. \eqref{genconditionr1} is always zero or negative, this inequality is always fulfilled whenever $r_1$ or $r_2$
are larger that $1/2$ and actually allows them to be smaller than this threshold.
This actually means that the purification protocol is able to convert the entanglement of
other  states into $\ket{\Psi^-}$. 
Symmetry arguments lead to the analogue condition for purifying $\ket{\Psi^+}$ with $r_3$ and $r_4$, namely
\begin{align}
  (2r_4-1)
  (1-2r_3)>-(2{\rm Im [r_{34}]})^2-(2{\rm Re}[r_{12}])^2.
  \label{genconditionr4}
\end{align}
Eqs. \eqref{genconditionr1} and \eqref{genconditionr4} are the main result of this work and 
define the condition to purify entanglement in the presented protocol. These conditions compared with \eqref{conditionr1} and \eqref{conditionr4}
relax the requirement for the diagonal parameters, because the right-hand side of the inequalities in \eqref{genconditionr1} and \eqref{genconditionr4} can be negative
and by thus $r_1, r_2, r_3$ and $r_4$ can be smaller than $1/2$. This clearly means that there exists states which do not have an overlap with a Bell state larger than $1/2$
but their ensemble can be still purified to $\ket{\Psi^-}$ or $\ket{\Psi^+}$. 
In the subsequent discussion, we investigate the inequalities of Eqs. \eqref{genconditionr1} 
and \eqref{genconditionr4}.

\begin{figure}[t]
 \includegraphics[width=.4\textwidth]{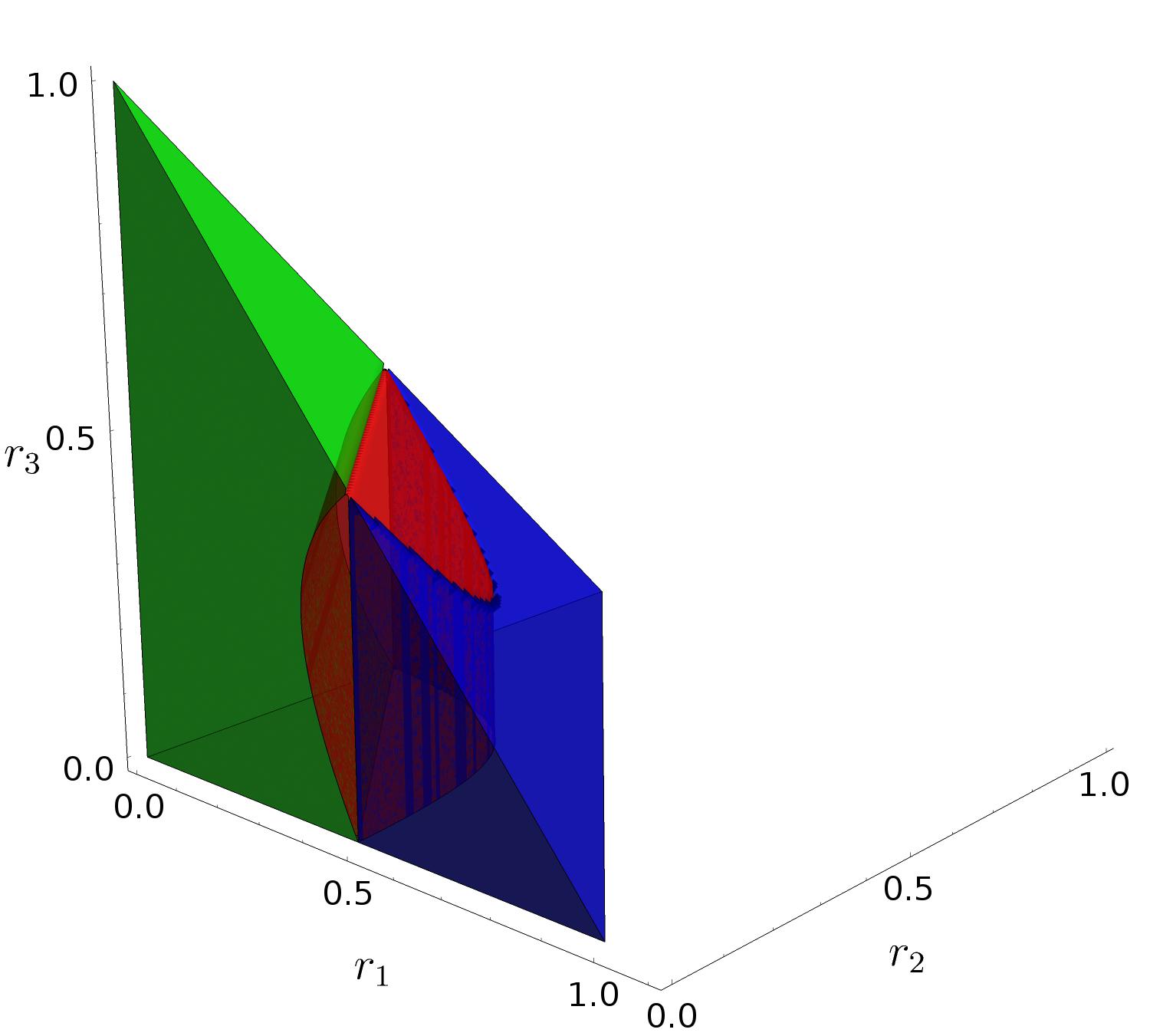}
 \caption{\label{figdeph1}
 Three dimensional space of the independent diagonal elements $(r_1,r_2,r_3)$. The other parameter
 $r_4$ is obtained form the trace condition ${\rm Tr}\rho=1$.
 In blue: the region defined by condition \eqref{genconditionr1} whose points lead to the fixed point $r_1=1$ and all others zero.
 In green: the region defined by condition \eqref{genconditionr4} that lead to the fixed point $r_4=1$. In red: all other
 conditions that do not lead to an entangled state. The initial coherence have the form in \eqref{dephasing1} with 
 $\eta_a=0.4$ and $\eta_b=0.1$.
 }
\end{figure}

In order to illustrate the difference with the previous section we
consider the following dependence of the coherences 
\begin{equation}
  r_{12}=\eta_a\sqrt{r_1r_2},\quad
  r_{34}=\eta_b\sqrt{r_3r_4},\quad \eta_a,\eta_b\in [0,1].
  \label{dephasing1}
\end{equation}
This case can be visualized in Fig. \ref{figdeph1} for the specific values of $\eta_a=0.4$ and $\eta_b=0.1$.
In the limit when $\eta_a,\eta_b\to1$, the area of red region tends to zero.

\begin{figure}[t]
 \includegraphics[width=.4\textwidth]{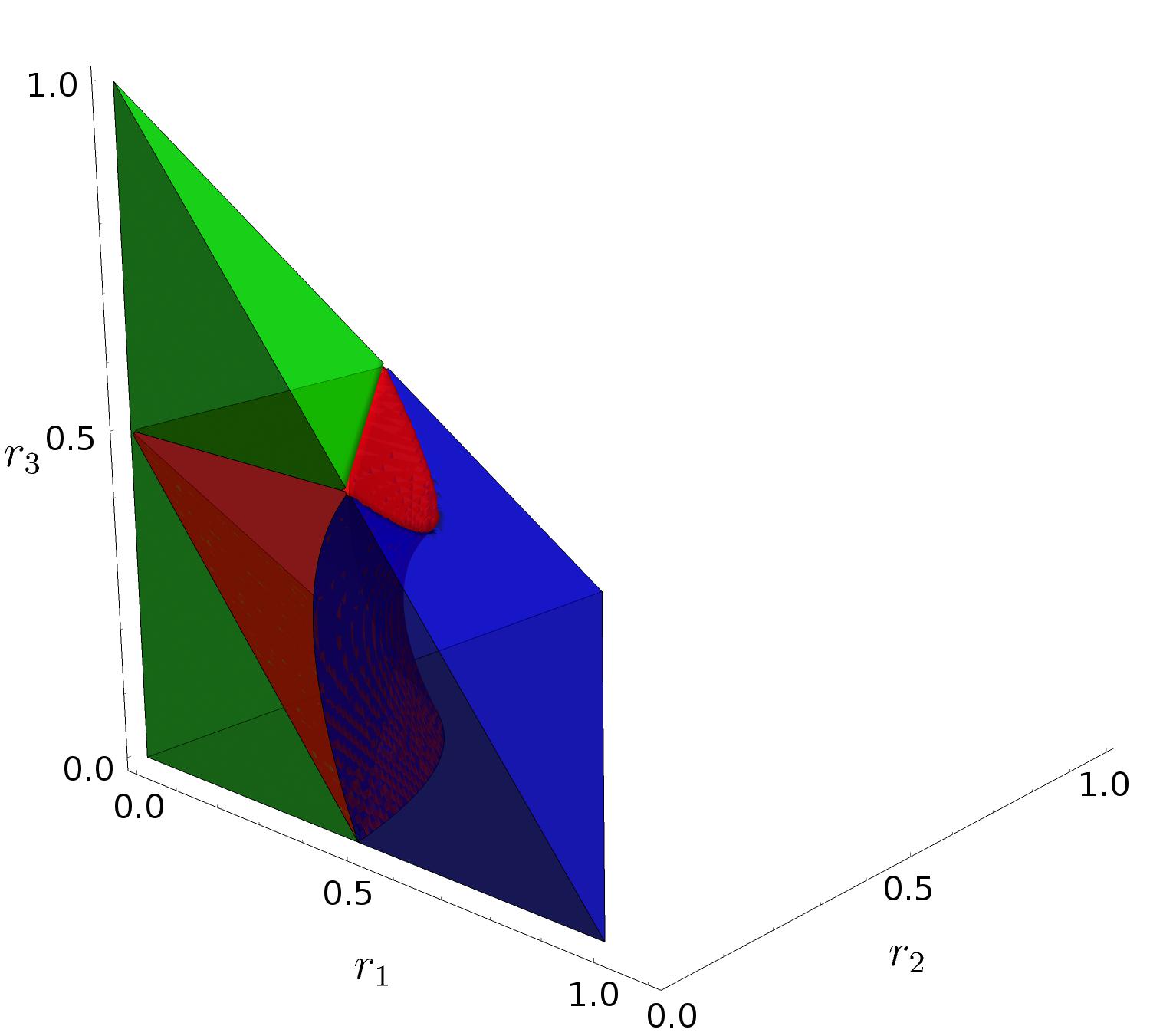}
 \caption{\label{figdeph2}
 Three dimensional space of the independent diagonal elements $(r_1,r_2,r_3)$. The other parameter
 $r_4$ is obtained form the trace condition ${\rm Tr}\rho=1$.
 In blue: the region defined by condition \eqref{genconditionr1} whose points lead to the fixed point $r_1=1$ and all others zero.
 In green: the region defined by condition \eqref{genconditionr4} that lead to the fixed point $r_4=1$. In red: all other
 conditions that do not lead to an entangled state. The initial coherence have the form in \eqref{dephasing2} with 
 $\eta_c=0.3$ and $\eta_d=0.5$.}
\end{figure}

Another possibility is the following behaviour of the coherences
\begin{equation}
  r_{12}=i\eta_c\sqrt{r_1r_2},\quad
  r_{34}=\eta_d\sqrt{r_3r_3},\quad \eta_c,\eta_d\in [0,1]
  \label{dephasing2}
\end{equation}
Figure \ref{figdeph2} shows the regions defined in this case for $\eta_c=0.3$ and $\eta_d=0.5$.
Also in this case, the limiting case  $\eta_c,\eta_d\to1$ corresponds to a vanishing area of the red region
(states whose fixed point is the total mixture).

These results demonstrate that the conditions on the diagonal parameters can be indeed relaxed.
Figures \ref{figdeph1} and \ref{figdeph2} show that, depending on the type of initial states,
the purifiable parameter space can be enlarged. Smaller values for the diagonal parameters
require higher absolute values for the coherences and meaning these state will be still entangled. 
This is of course expected, as one cannot create entanglement with local operations.

\section{Examples}
\label{Examples}
In this section we provide examples of states that fulfill the condition in Eq. \eqref{genconditionr1}.
The examples show how our protocol is able to purify initial entangled states with low fidelity with respect 
to Bell states.


{\it Example 1.} Consider the parameter $x\in (0.5,1]$ determining  the state 
\begin{align}
  \rho_1&=x\ketbra{\Upsilon_{\rm ent}}{\Upsilon_{\rm ent}}+(1-x)\ketbra{\Phi^+}{\Phi^+}
  \nonumber\\
  \ket{\Upsilon_{\rm ent}}&=\tfrac{1}{\sqrt2}\left(\ket{\Psi^-}+ i\ket{\Phi^-}\right).
  \label{}
\end{align}
The state $\rho_1$ has no fidelity larger than one half with respect to any of the 
four Bell states. However, it has fidelity $x>1/2$ with respect to the maximally entangled
state $\ket{\Upsilon_{\rm ent}}$. Taking into account that in this example 
$r_1=r_2=i r_{12}=x/2$ and $r_{34}=0$ one can find that after one iteration of the map
\begin{align*}
  r_1'=\frac{x^2}{x^2+(1-x)^2}
  \label{}
\end{align*}
which is always larger than $x$ for $x\in (0.5,1)$. This is an example 
of an initial state where the
fidelities with respect to any of the four Bell states is smaller than one half
and still the state $\ket{\Psi^-}$ can be purified.  
Of course, here the fidelity with respect to the maximally entangled state $\ket{\Upsilon _{\rm ent}}$
is being exploited by the protocol. In contrast, this feature would  be lost with other
known protocols \cite{Bennett1,Deutsch} that use random unitary gates to bring a state into
a Bell diagonal form. This procedure would destroy the entanglement of $\rho_1$.

{\it Example 2.} Now take the parameter $c\in (0,0.5]$ that parametrizes the state
\begin{align}
  \rho_{2}&=c\ketbra{\Psi^-}{\Psi^-}+(1-c)\ketbra{\Upsilon_{\rm sep}}{\Upsilon_{\rm sep}}
  \nonumber\\
  \ket{\Upsilon_{\rm sep}}&=\frac{1}{\sqrt2}
  \left(\ket{\Phi^+}+\ket{\Psi^+}\right).
  \label{}
\end{align}
In this example $r_1=c$ and $r_3=r_4=r_{34}=(1-c)/2$. After a single iteration of the protocol one finds that the fidelity with respect to
$\ket{\Psi^+}$ takes the value
\begin{align*}
  r_1'=\frac{c^2}{2c^2+(1-2c)-(1-c)^2}=1.
  \label{}
\end{align*}
This means that no matter the value of $c$, as long it is not zero, the state can be purified
to $\ket{\Psi^-}$ in just one iteration of the protocol. 
The success probability is $c^2/2$ and therefore
the smaller the value of $c$ the less probable is the occurrence of an event that generates $\ket{\Psi^-}$.
However, the fact that perfect purification is achieved in just one step makes
this protocol with this initial state an attractive candidate. 
Typically the overall success probability
scales exponentially with the number of steps \cite{DürBriegel} and the convergence is formally achieved 
in an asymptotic way.

Interestingly, state $\rho_2$ has overlap smaller than one half with any maximally entangled pure state. 
In order to proof this statement, let us consider the general form of these type of states in the Bell basis.
First, we recall that the concurrence \cite{Wootters} is a measure of entanglement
which takes unit value for maximally entangled pure states. 
For a pure state \ket{\Psi} it is defined as
\begin{align}
  C(\ket\Psi)=|\bra{\Psi}\sigma_y\otimes\sigma_y\ket{\Psi}^\ast|,
  \label{concurrencepure}
\end{align}
with the Pauli matrix $\sigma_y$ and
where $\ket\Psi^\ast$ is the complex conjugated vector of $\ket\Psi$. 
The Bell states are eigenstates of
the operator in Eq. \eqref{concurrencepure}, namely 
$\sigma_y\otimes\sigma_y\ket{\Psi^\pm}=\pm\ket{\Psi^\pm}$ and
$\sigma_y\otimes\sigma_y\ket{\Phi^\pm}=\mp\ket{\Phi^\pm}$. With this one can note that, up to a global phase, 
any maximally entangled pure state has the form
\begin{align}
  \ket\Psi=a_-\ket{\Phi^-}+ia_+\ket{\Phi^+}+ib_-\ket{\Psi^-}+b_+\ket{\Psi^+},
  \label{}
\end{align}
with coefficients $a_\pm,b_\pm\in\mathbb{R}$ and satisfying  
the normalization condition  $\braket{\Psi}{\Psi}=1$.
Then, by introducing $y=(a_+^2+b_+^2)/2$, the fidelity of $\rho_2$ with respect to any maximally 
entangled pure state takes the form
\begin{align}
  F_{\rho_2}=\bra{\Psi}\rho_2\ket{\Psi}&= 
  c b_-^2+(1-c)y\\
&\le c \left(1-y\right)+(1-c)y\le\frac{1}{2}.\nonumber
  \label{}
\end{align}
The first inequality takes into account the normalization conditions of $\ket\Psi$.
The last inequality is valid provided both $c$ and $y$ are positive numbers less than a half, which is true
by our choice of $c$ and by the definition of $y$. 

Finally, let us point out that the parameter $c$ is actually the concurrence \cite{Wootters} of the state $\rho_2$. 
Indeed, it is not hard to realize that the matrix 
  $\rho_2 (\sigma_y\otimes\sigma_y)\rho_2^\ast(\sigma_y\otimes\sigma_y)$
  has only one eigenvalue given by $c^2$ and therefore, according to Ref. \cite{Wootters}, the concurrence is $c$.

\section{Conclusions}
\label{conclusions}

We have studied the convergence of arbitrary input two-qubit states,
under the influence of the entanglement purification protocol presented 
in Sec. \ref{PurifyP} and first introduced in Ref. \cite{Bernad}. 
Our protocol was introduced in the context of a multiphoton-assisted quantum repeater,
where a two-qubit quantum operation can be more efficiently realized than a controlled-NOT gate. 

In a first approach, we have studied in detail the convergence 
in a subset of density matrices which is left invariant by the protocol.  
This subset is characterized by seven real parameters, three of which are diagonal elements of the density
matrix in the Bell basis. These diagonal elements do not depend on the coherences after the iteration of the protocol.
Therefore, we have based our proof on the work of C. Macchiavello \cite{Macchiavello}. 
In order to show that the off-diagonal elements monotonically decrease, 
we have introduced a quadratic form which decreases under each iteration of the map.
Combining these results we were able to define those conditions for the initial states which lead 
to the purification of two Bell states. 
We have shown that these conditions can be generalized to arbitrary density 
matrices with $15$ real parameters.
This contrasts with the two seminal entanglement purification protocols \cite{Bennett1,Deutsch} which consider
density matrices characterized by one or three real parameters. Random unitary rotations are required in those cases 
to bring any state into a Bell diagonal state. This step in the original protocols  may 
destroy useful entanglement. The fact that our protocol does not require such step is the 
reason why we obtain less constrained conditions for purifiable initial states. 

Exploiting these new findings, we have shown that our protocol allows purification of entangled states having
an overlap less than one half with any Bell state. Some states might even have fidelity less than one half
with respect to any maximally entangled pure state. We have shown a class of these type of states that can be converted
into a Bell state in just one step of the purification protocol. The states are parametrized by their concurrence
$c$ and the probability of success is given by $c^2/2$. This feature is interesting, as the resources of
entanglement purification scale exponentially with the number of states. Therefore, our result could  initiate the
study of other type of local two-qubit operations which might lead to an optimal entanglement purification for 
specific type of states. Provided that the source of the entangled pairs is well under control, the engineering
of an efficient local two-qubit operation might be more resourceful than the application of a controlled-NOT gate.

\begin{acknowledgments}
This work is supported by the BMBF project Q.com.
\end{acknowledgments}

\appendix

\section{Fixed point analysis of $X$ states}
\label{App}

In order to show that there are no other stable fixed points in regions 
defined by the conditions \eqref{conditionr1} and \eqref{conditionr4}, 
we have numerically investigated the stability of the fixed points of the map \eqref{map}. 
This is accomplished by studying the Jacobian matrix with entries $J_{ij}$. 
The entries can be defined defined with the help of the 
function $\boldsymbol f: \mathbb{R}^8 \to \mathbb{R}^8$ as $J_{i,j}=\frac{\partial f_i}{\partial x_j}$,
where each $x_i$ is a component of the  eight dimensional vector
$\boldsymbol v=(r_1,r_2,r_3,r_4,\mathrm{Re}[r_{14}],\mathrm{Im}[r_{14}],\mathrm{Re}[r_{23}],\mathrm{Im}[r_{23}])$.
The introduced function fulfills the relation $\boldsymbol v'=\boldsymbol f(\boldsymbol v)$
where the primed variables are defined in the non-linear map \eqref{map}. Fixed points are obtained by
solving the equation $\boldsymbol v_\star=\boldsymbol f(\boldsymbol v_\star)$ and their stability is specified
with the eigenvalues of $J|_{\boldsymbol v=\boldsymbol v_\star}$. If all eigenvalues have absolute value smaller
than one, the point is stable. If the absolute value of any eigenvalue is larger than one, the fixed point is
unstable.


Table \ref{table1} shows that there are three stable  fixed points, of which two represent the cases when we purify the states $\ket{\Psi^-}$ and $\ket{\Psi^+}$ and the other
when we converge towards the maximally mixed state. 
We have found that when considering initial density matrices with one of the parameters $r_i=0$ and the rest 
$r_j,r_k,r_l \in [0,0.5]$ the iteration of the protocol can
lead to fixed points of at least second order, meaning that they are fixed points of the 
second iteration of the function $\boldsymbol f$, namely 
$\boldsymbol v_{\star\star}=\boldsymbol f(\boldsymbol f(\boldsymbol v_{\star\star}))$. 
It also can happen that with right initial conditions the map stays or cycles 
around unstable one- or multi-period fixed points.

\begin{table}[h]
\begin{center}
  \begin{tabular}{ll}
    \begin{tabular}{l}
     
     \,\,\,\,\,\,\,\, Fixed points \\
     $(r_1,r_2,r_3,r_4,r_{14},r_{23})$
    \end{tabular}
   
    & \,\,\,\,\, Stability  
    \\
\hline\noalign{\bigskip}
   \,\,\,\,\, $(1,0,0,0,0,0)\,\,\,\,\,\,\,\,$  & \,\,\,\,\,\,\,\,\, stable 
    \\
\noalign{\smallskip}
   \,\,\,\,\, $(0,0,0,1,0,0)\,\,\,\,\,\,\,\,$  & \,\,\,\,\,\,\,\,\, stable 
    \\
\noalign{\smallskip}
   \,\,\,\,\, $(0.1409, 0.2344, 0.1245, 0.5,0,0)\,\,\,\,\,\,\,\,$  & \,\,\,\,\,\,\,\,\, unstable 
    \\
\noalign{\smallskip}
   \,\,\,\,\, $(0.5, 0.1245, 0.2344, 0.1409,0,0)\,\,\,\,\,\,\,\,$  & \,\,\,\,\,\,\,\,\, unstable
       \\
\noalign{\smallskip}
   \,\,\,\,\, $(0.25, 0.25, 0.25, 0.25,0,0)\,\,\,\,\,\,\,\,$  & \,\,\,\,\,\,\,\,\, stable 
       \\
\noalign{\smallskip}
   \,\,\,\,\, $(0.25, 0.25, 0.25, 0.25,0.25,-0.25)\,\,\,\,\,\,\,\,$  & \,\,\,\,\,\,\,\,\, unstable 
         \\
\noalign{\smallskip}
   \,\,\,\,\, $(0.25, 0.25, 0.25, 0.25,0.25,0.25)\,\,\,\,\,\,\,\,$  & \,\,\,\,\,\,\,\,\, unstable 
   \\
\noalign{\smallskip}
   \,\,\,\,\, $(0.5, 0, 0, 0.5,0,0)\,\,\,\,\,\,\,\,$  & \,\,\,\,\,\,\,\,\, unstable 
         \\
\noalign{\smallskip}
   \,\,\,\,\, $(0.5, 0, 0, 0.5,0.5,0)\,\,\,\,\,\,\,\,$  & \,\,\,\,\,\,\,\,\, unstable 
  \end{tabular}
\end{center}
\caption{ 
\label{table1}
List of fixed points of the map \eqref{map} with their respective stability. Two fixed points are given with 4-digit precision.
}
\end{table}


\begin{thebibliography}{99}
%
\bibitem{Bennett1} C. H. Bennett, G. Brassard, S. Popescu, B. Schumacher, J. A. Smolin, and W. K. Wootters, Phys. Rev. Lett. {\bf 76}, 722 (1996); {\bf 78}, 2031 (1997).
%
\bibitem{Deutsch} D. Deutsch, A. Ekert, R. Jozsa, C. Macchiavello, S. Popescu, and A. Sanpera, Phys. Rev. Lett. {\bf 77}, 2818 (1996); {\bf 80}, 2022 (1998).
%
\bibitem{Bennett2} C. H. Bennett, D. P. DiVincenzo, J. A. Smolin, and W.K. Wootters, Phys. Rev. A {\bf 54}, 3824 (1996).
%
\bibitem{DürBriegel} W. D\"ur and H. J. Briegel, Rep. Prog. Phys. {\bf 70}, 1381 (2007).
%
\bibitem{Briegel98} H.-J. Briegel, W. D\"ur, J. I. Cirac, and P. Zoller, Phys. Rev. Lett. {\bf 81}, 5932 (1998).
%
\bibitem{Werner} R. F. Werner, Phys. Rev. A {\bf 40}, 4277 (1989).
%
\bibitem{Macchiavello} C. Macchiavello, Phys. Lett. A {\bf 246}, 345 (1998).
%
\bibitem{Bernad} J. Z. Bern\'ad, J. M. Torres, L. Kunz, and G. Alber, Phys. Rev. A {\bf 93}, 032317 (2016).
%
\bibitem{Haroche} J. M. Raimond, M. Brune, and S. Haroche, Rev. Mod. Phys. {\bf 73}, 565 (2001).
%
\bibitem{Wootters} W. K. Wootters, Phys. Rev. Lett. {\bf 80}, 2245 (1998).
%
\end{thebibliography}
\end{document}